\documentclass[onecolumn]{ceurart}

\usepackage{listings}
\usepackage{subcaption}
\begin{document}
    
    \copyrightyear{2025}
    \copyrightclause{Copyright for this paper by its authors.
        Use permitted under Creative Commons License Attribution 4.0
        International (CC BY 4.0).}
    
    \conference{IntRS'25: Joint Workshop on Interfaces and Human Decision Making for Recommender Systems, September 22, 2025, Prague}
    
    \title{Towards LLM-Based Usability Analysis for Recommender User Interfaces}
    
    \author[1]{Sebastian Lubos}[%
    orcid=0000-0002-5024-3786,
    email=sebastian.lubos@tugraz.at,
    url=https://ase.sai.tugraz.at/,
    ]
    \cormark[1]
    \address[1]{Graz University of Technology, Inffeldgasse 16b, Graz, 8010, Austria}

    \author[1]{Alexander Felfernig}[%
    orcid=0000-0003-0108-3146,
    email=alexander.felfernig@tugraz.at,
    url=https://ase.sai.tugraz.at/,
    ]

    \author[1]{Damian Garber}[%
    orcid=0009-0005-0993-0911,
    email=damian.garber@tugraz.at,
    url=https://ase.sai.tugraz.at/,
    ]
    
    \author[1]{Viet-Man Le}[%
    orcid=0000-0001-5778-975X,
    email=v.m.le@tugraz.at,
    url=https://ase.sai.tugraz.at/,
    ]

    \author[1]{Thi Ngoc Trang Tran}[%
    orcid=0000-0002-3550-8352,
    email=ttrang@ist.tugraz.at,
    url=https://ase.sai.tugraz.at/,
    ]
    
    \cortext[1]{Corresponding author.}

\begin{abstract}
Usability is a key factor in the effectiveness of recommender systems. However, the analysis of user interfaces is a time-consuming process that requires expertise. Recent advances in \textit{multimodal large language models (LLMs)} offer promising opportunities to automate such evaluations. In this work, we explore the potential of multimodal LLMs to assess the usability of recommender system interfaces by considering a variety of publicly available systems as examples. We take user interface screenshots from multiple of these recommender platforms to cover both \textit{preference elicitation} and \textit{recommendation presentation} scenarios. An LLM is instructed to analyze these interfaces with regard to different usability criteria and provide explanatory feedback. Our evaluation demonstrates how LLMs can support heuristic-style usability assessments at scale to support the improvement of user experience.
\end{abstract}


\maketitle

\section{Introduction}\label{sec:introduction} 
Recommender systems are a central component of many digital platforms, where they provide personalized item suggestions to help users navigate large sets of options~\cite{ricci2022recommender}. While the quality of the underlying recommendation algorithm is important~\cite{zangerle2022evaluating}, the overall effectiveness of a recommender system also depends on how well users can interact with the interface. Usability and user experience play a key role in enabling users to express preferences, interpret recommendations, and make informed choices~\cite{pu2012evaluating,knijnenburg2012explaining}. Even highly accurate recommendations may fail to deliver value if the interface is difficult to navigate or lacks transparency.

Traditional usability evaluation methods include \textit{usability testing} with real users~\cite{hass2019practical} and \textit{expert inspections} based on heuristic principles~\cite{holingsed2007usability}. While these methods are effective, they are time-consuming and require expert involvement. General usability guidelines, such as \textit{Nielsen’s heuristics}~\cite{nielsen1994enhancing} offer structured support, and recommender-specific frameworks further improve contextual relevance~\cite{pu2011user}. Nevertheless, usability assessments remain resource-intensive and are thus rarely applied across platforms.

To reduce this effort, automated solutions such as rule-based tools and heuristic checkers have been proposed~\cite{namoun2021review,castro2022automated}. However, these often cover only limited usability dimensions and struggle with subjective or context-specific issues~\cite{kuric2025systematic}. More recently, \textit{multimodal large language models (LLMs)} that can process both visual and textual inputs have emerged ~\cite{yin2024survey}. Early studies demonstrate their ability to identify usability issues in design mockups~\cite{duan2024generating} and mobile interfaces~\cite{pourasad2025does}, although expert validation remains necessary. Initial research on the alignment between LLM-based analyses and expert assessments reports promising accuracy in different scenarios~\cite{zhong2025synthetic,lubos2025towards}, but more studies are needed to confirm these results. 

While these approaches address general usability, they have not yet been applied to the specific challenges of recommender system interfaces, such as explainability, feedback mechanisms, and preference elicitation workflows, which are central to the user experience in this context. In this work, we explore how a multimodal LLM can help to analyze the usability of ten publicly available recommender interfaces based on explicitly defined criteria. We review the analysis results to highlight the feasibility and benefits of automated usability analysis for recommender interfaces and outline directions for future research.

The paper is organized as follows: Section~\ref{sec:usability_analysis} describes the experimental setup and implementation details. Section~\ref{sec:results} presents the results. Section~\ref{sec:discussion} discusses implications and future work. Finally, the paper is concluded in Section \ref{sec:conclusions}.

\section{Usability Analysis of Recommender Interfaces}\label{sec:usability_analysis} 
A lot of focus in recommender system research is put on the accuracy of algorithms and personalization strategies \cite{zangerle2022evaluating,gunawardana2022evaluating}. However, the quality of recommender user interfaces plays an equally important role in shaping the overall user experience. This experience can be assessed through \textit{usability analysis} that is concerned with the aspect of how effectively users can navigate, interpret, and interact with different parts of the recommender system \cite{pu2012evaluating,knijnenburg2012explaining}. These include possibilities to express explicit preferences, understand why items are recommended, review recommended items, and provide feedback. 

The following sections outline our considered recommender scenarios, usability criteria, and describe the LLM-based analysis in detail.

\subsection{Recommender Scenarios}
To explore the LLM-based usability analysis across a diverse set of recommender interfaces, we selected ten publicly accessible platforms from various item domains, which are summarized in Table \ref{tbl:scenarios}. These systems vary in layout complexity, interaction mechanisms, and types of recommendations. This way, the automated usability analysis could be reviewed for varying contexts to get an impression about its generalizability. Each platform was assessed in two typical usage scenarios: (i) \textit{preference elicitation}, and (ii) \textit{recommendation presentation}. 

\begin{table}[h]
    \small
    \centering
    \begin{tabular}{|l|l|l|}
    \hline
    \textbf{Platform} & \textbf{Item Domain} & \textbf{URL} \\
    \hline
        Amazon & E-commerce & \url{https://www.amazon.com} \\
        Goodreads & Books & \url{https://www.goodreads.com} \\
        Google News & News Articles & \url{https://news.google.com} \\
        KaptnCook & Recipes & \url{https://www.kaptncook.com} \\
        Last.fm & Music & \url{https://www.last.fm} \\
        Netflix & Movies \& TV Shows & \url{https://www.netflix.com} \\
        Pinterest & Visual Content & \url{https://www.pinterest.com} \\
        Spotify & Music & \url{https://open.spotify.com} \\
        Steam & Video Games & \url{https://store.steampowered.com} \\
        YouTube & Videos & \url{https://www.youtube.com} \\
    \hline
    \end{tabular}
    \caption{Overview of selected recommender platforms for usability analysis.}
    \label{tbl:scenarios}
    \vspace{-8pt}
\end{table}

To have a comparable situation for each platform, we considered a new user scenario of a user interacting with the recommender for the first time. For this purpose, we used a desktop browser in incognito mode to avoid personalization effects and simulate a first-time user experience.\footnote{A new user account was registered if needed to use the application.} For each platform, we captured a representative screenshot\footnote{Screenshots were taken at full resolution and included the visible viewport with relevant UI context (e.g., navigation bars, filters, recommendation labels).} for the usage scenarios. Depending on the platform, the \textit{preference elicitation} either showed the default onboarding screens or initial filter options. The \textit{recommendation presentation} showed either the main homepage (dashboard) or a detail page including recommended items.

To ensure comparability across platforms while accounting for platform-specific interaction patterns, we defined explicit user tasks for each platform. This allowed us to maintain a consistent evaluation structure while respecting the nuances of individual interfaces. The tasks for both scenarios and all platforms are presented in Table~\ref{tbl:tasks}. They were used to select the screenshots and provide contextual information to the LLM during the usability analysis.

\begin{table}[h]
    \small
    \centering
    \begin{tabular}{|p{2.0cm}|p{5.4cm}|p{5.8cm}|}
    \hline
    \textbf{Platform} & \textbf{Preference Elicitation Task} & \textbf{Recommendation Presentation Task} \\
    \hline
    Amazon & Search for “Bluetooth headphones” and interact with product listings to express shopping intent. & Review the related product recommendations shown on the product or search results page. \\
    \hline
    Goodreads & Rate previously read books as part of the onboarding process to express reading preferences. & Review the initial book recommendations generated based on ratings. \\
    \hline
    Google News & Select preferred news topics or regions during setup to tailor content delivery. & Review the personalized news feed on the dashboard. \\
    \hline
    KaptnCook & Indicate disliked ingredients during the initial setup to personalize meal suggestions. & Review the list of recommended recipes based on stated preferences. \\
    \hline
    Last.fm & Choose a trending artist to indicate music preferences. & Review the list of recommended tracks or artists based on the selected input. \\
    \hline
    Netflix & Select at least three preferred titles during the onboarding setup to express content preferences. & Review the personalized dashboard with recommended movies and shows. \\
    \hline
    Pinterest & Select inspirational images reflecting personal interests during the onboarding process. & Review the personalized feed with recommended visual content. \\
    \hline
    Spotify & Add songs to playlists based on initial recommendations to express musical preferences. & Review the homepage or dashboard with recommended tracks. \\
    \hline
    Steam & Apply filters (e.g., "Indie" genre) while browsing to express game preferences. & Review the recommended games displayed on the store homepage or Discovery Queue. \\
    \hline
    YouTube & Search for and watch a video on tennis serve drills to signal viewing preferences. & Review the recommended videos shown after watching the selected content. \\
    \hline
    \end{tabular}
    \caption{Standardized user task descriptions for preference elicitation and recommendation presentation.}
    \label{tbl:tasks}
    \vspace{-16pt}
\end{table}

\subsection{Usability Criteria}
To analyze the usability of the recommender system interfaces in a structured way, we defined a set of criteria that are shown in Table \ref{tbl:criteria}. These are based on general established usability principles \cite{nielsen1994enhancing} and user-centric evaluation metrics for recommender systems \cite{pu2011user}. We defined and adapted these criteria to cover general interface qualities and scenario-specific aspects of \textit{preference elicitation} and \textit{recommendation presentation}. Our goal was to assess whether recommender interfaces support users in understanding, influencing, and responding to recommendations.

\begin{table}[h]
    \small
    \centering
    \begin{tabular}{|p{3.0cm}|p{10.7cm}|}
    \hline
    \textbf{Category} & \textbf{Usability Criterion} \\
    \hline
    \multirow{4}{*}{\shortstack[l]{General\\(Both Scenarios)}} 
    & \textit{G1.} Is the layout clear and visually structured? \\
    & \textit{G2.} Are interactive elements (e.g., buttons, icons) clearly recognizable? \\
    & \textit{G3.} Is the amount of information per item appropriate and helpful? \\
    & \textit{G4.} Are interface elements used consistently (e.g., icons, labels, colors)? \\
    \hline
    \multirow{3}{*}{Preference Elicitation} 
    & \textit{P1.} Can users explicitly express preferences (e.g., ratings, likes, categories)? \\
    & \textit{P2.} Is there transparency about how input affects recommendations? \\
    & \textit{P3.} Do users have control and flexibility (e.g., skip, edit, undo inputs)? \\
    \hline
    \multirow{4}{*}{\shortstack[l]{Recommendation\\Presentation}} 
    & \textit{R1.} Are recommendations clearly labeled as such? \\
    & \textit{R2.} Are different types of recommendations distinguishable (e.g., "Because you liked...")? \\
    & \textit{R3.} Are there explanations for why items are recommended? \\
    & \textit{R4.} Can users interact with recommendations (e.g., rate, hide, save)? \\
    \hline
    \end{tabular}
    \caption{Usability criteria used to analyze recommender interfaces.}
    \label{tbl:criteria}
    \vspace{-12pt}
\end{table}

\subsection{LLM-based Usability Analysis}

We used the \textit{gemini-2.5-flash} model by Google for the LLM-based usability analysis ~\cite{geminiteam2024gemini}. This model was designed to process textual and visual inputs, which makes it suitable for our experiments. We instructed the model in Python using the Gemini Developer API.\footnote{\url{https://ai.google.dev/gemini-api/docs}} To improve the reproducibility of results, we set the \textit{temperature} to $0.0$ for more deterministic LLM output. 

\begin{figure}[h]
    \centering
    \includegraphics[width=0.9\linewidth]{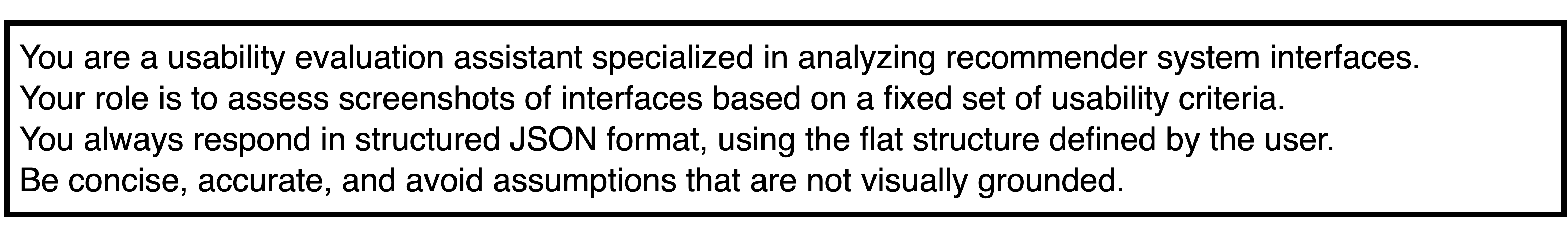}
    \vspace{-8pt}
    \caption{System prompt for usability analysis.}
    \label{fig:system_prompt}
    \vspace{-16pt}
\end{figure}

For the analysis task, the recommender interface screenshot was provided as context, together with a high-level description of the platform, the considered usage scenario (\textit{preference elicitation} or \textit{recommendation presentation}), and an explicit user task (see Table~\ref{tbl:tasks}). To define the role and boundaries of the LLM in this scenario, we used the \textit{system prompt} shown in Figure~\ref{fig:system_prompt}.

\begin{figure}[h]
    \centering
    \includegraphics[width=0.9\linewidth]{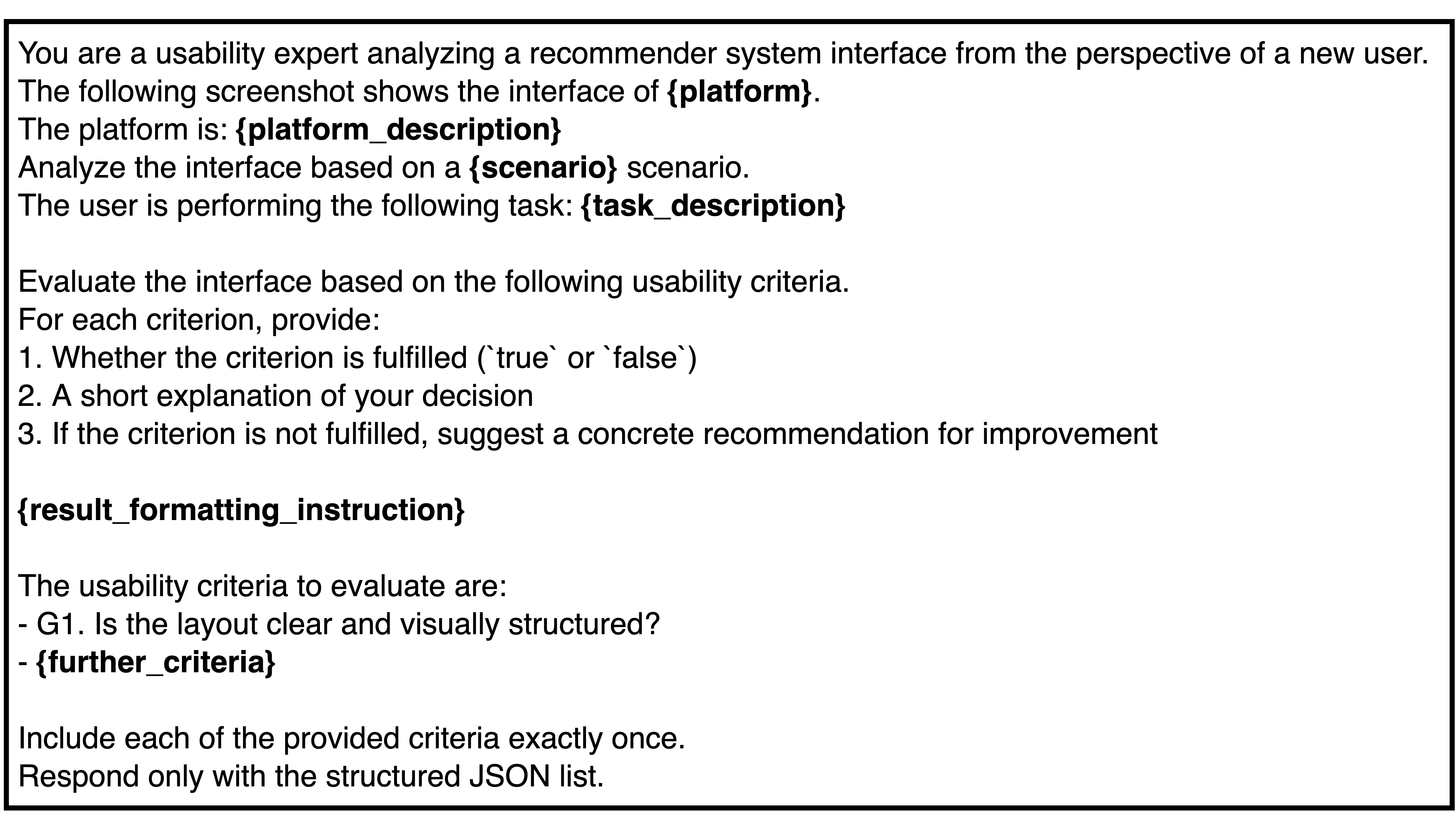}
    \vspace{-8pt}
    \caption{User prompt template for usability analysis. Variables are highlighted in braces (\textbf{\{\}}).}
    \label{fig:user_prompt}
    \vspace{-16pt}
\end{figure}

Figure~\ref{fig:user_prompt} shows the \textit{user prompt} template to instruct the LLM for the explicit analysis of the provided screenshot. It describes the expected output, which includes a binary evaluation of whether the usability criterion is fulfilled or not, an explanation to justify its decision, and a suggestion on how to improve this usability aspect in case the criterion was not fulfilled. The respective list of usability criteria for each scenario (see Table~\ref{tbl:criteria}) was provided in the prompt.

\section{Results}\label{sec:results} 
We analyzed $10$ platforms across $2$ usage scenarios each, and considered $11$ different usability criteria. This resulted in $150$ individual assessments (respecting partly scenario-specific criteria, see Table~\ref{tbl:criteria}). The LLM completed the analysis in approximately $216$ seconds.

\begin{figure}[h]
\centering
\includegraphics[width=0.7\linewidth]{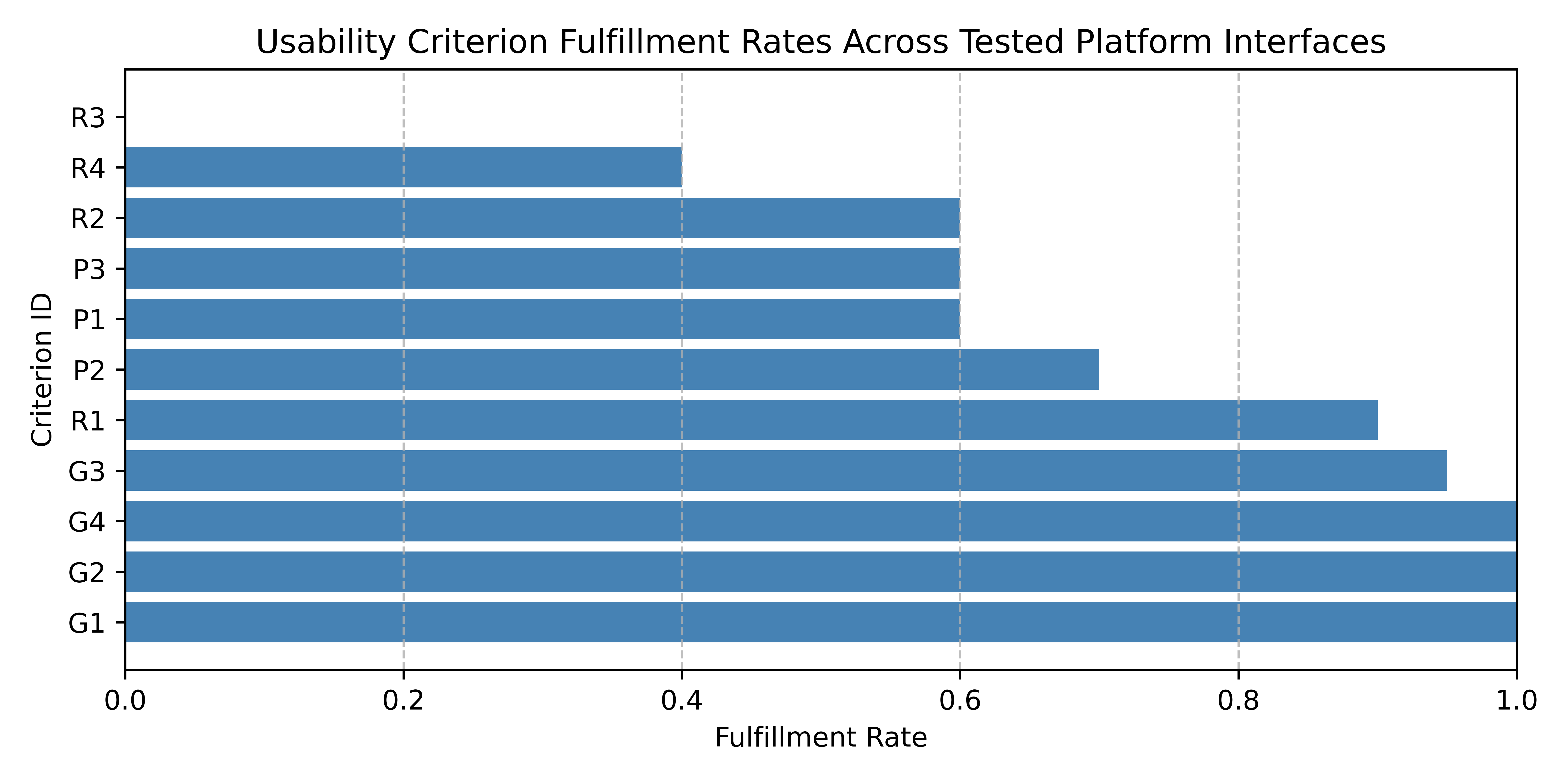}
\caption{Fulfillment rates for each usability criterion across all platforms. The horizontal bars indicate the proportion of platforms for which each criterion was considered fulfilled by the LLM-based analysis.}
\vspace{-8pt}
\label{fig:fulfillment}
\vspace{-4pt}
\end{figure}

Figure~\ref{fig:fulfillment} summarizes the fulfillment rates for each usability criterion across all evaluated platforms. The results show that general interface design aspects, such as clear layout (G1), recognizable interactive elements (G2), appropriate information density (G3), and consistent visual styling (G4), are well-supported on most platforms. This suggests that these systems follow established design conventions and offer a solid baseline for usability. This outcome is expected, given that the evaluated platforms are widely used and likely to have invested greatly in interface design and user experience.

In contrast, the recommender-specific criteria show a more diverse picture. Particularly, the presence of explanations (R3) and interactive feedback options (R4) was less frequently fulfilled. The same holds for transparency of underlying algorithms (R2/P2), explicit feedback options (P1), and flexibility of interaction (P3). This suggests that while basic UI design is generally strong, many platforms lack mechanisms to help users understand and influence recommendation behavior. These findings point to a gap in explainability and user control in many ``\textit{black-box}'' recommender settings.

\begin{figure}[h]
\centering
\includegraphics[width=0.7\linewidth]{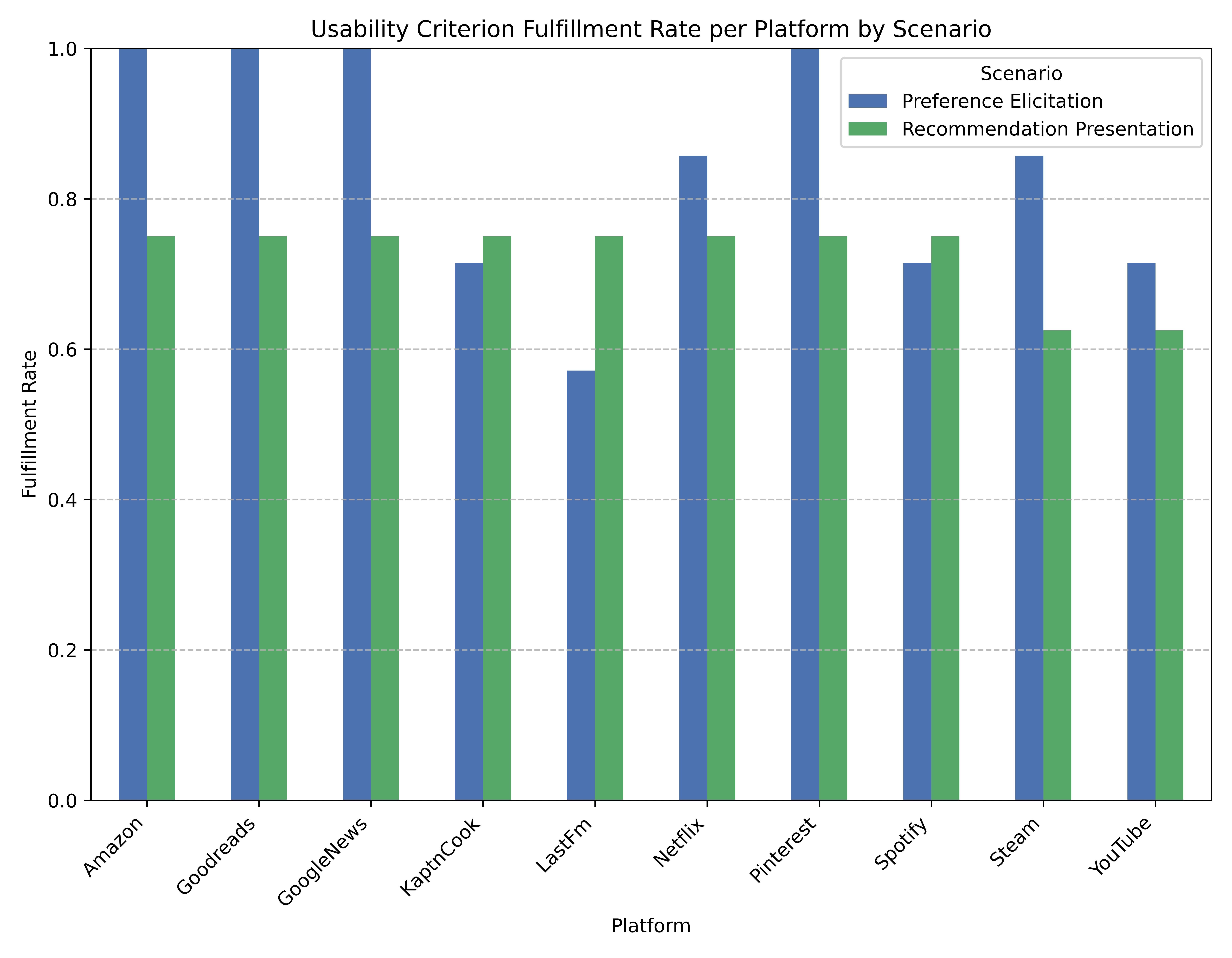}
\vspace{-12pt}
\caption{Fulfillment rates of usability criteria per platform and usage scenario.}
\label{fig:platform_fulfillment}
\vspace{-12pt}
\end{figure}

Figure~\ref{fig:platform_fulfillment} shows the average fulfillment rates for each platform, separated by usage scenario. The results indicate that most platforms performed slightly better in the \textit{preference elicitation} scenario, where users explicitly express interests or filters. In contrast, in the \textit{recommendation presentation} scenario, where users review recommended items, more usability issues were revealed. Still, the overall high fulfillment rates are expected as all platforms are widely used.

The quantitative results offer an initial overview of the LLM-based usability analysis. To better understand the practical value of the approach, we examined individual outputs in detail. The following examples show how different criteria were analyzed across platforms.

\begin{figure}[h]
\centering
\includegraphics[width=0.9\linewidth]{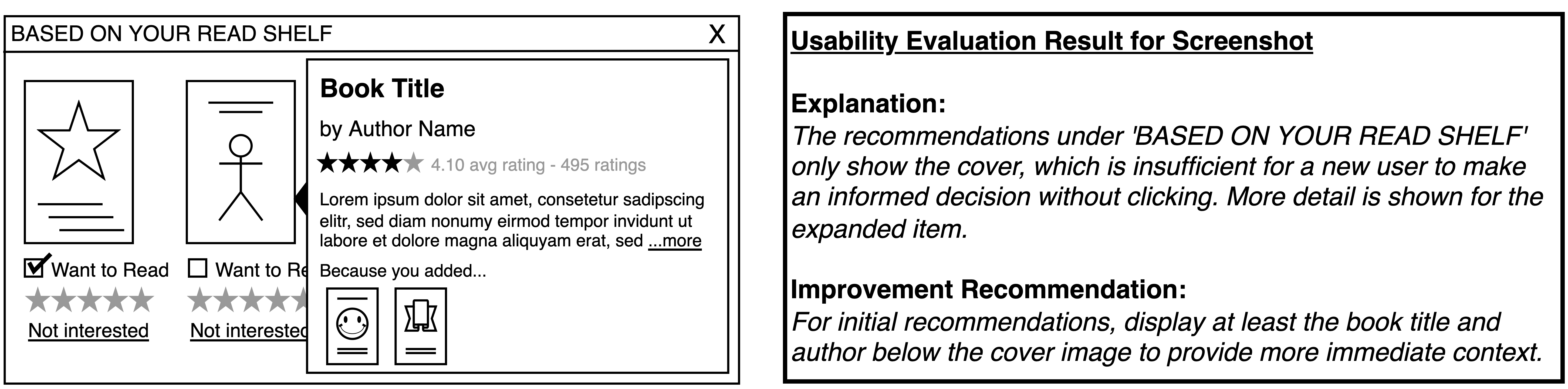}
\caption{Example usability analysis result for \textit{Goodreads} on the \textit{unfulfilled information per item (G3)} criterion, with corresponding explanations and improvement suggestions. The sketch illustrates the analyzed UI.
}
\label{fig:goodreads_g3}
\vspace{-16pt}
\end{figure}

The analysis revealed only one unfulfilled general usability criterion related to the amount of information shown per item (G3) on the \textit{Goodreads} platform. Figure~\ref{fig:goodreads_g3} presents the corresponding LLM explanation and improvement suggestion, alongside a UI sketch that highlights the issue.\footnote{To avoid potential legal issues with publishing proprietary platform screenshots, we redrew the relevant parts as simplified UI sketches for presentation in this paper. Importantly, all analyses were conducted using the original screenshots. The sketches only serve as substitutes for publication and preserve the necessary information for understanding the reported aspects. This procedure was applied to all examples shown. For the original screen design, we refer to the respective platform web pages (see Table~\ref{tbl:scenarios}).} In this case, the LLM makes a valid point: when book covers are small and titles are unreadable, it is difficult for users to get a first impression of the item without reviewing its details. The suggested improvement is appropriate and addresses this usability gap effectively.

\begin{figure}[h]
    \centering
    \begin{subfigure}[b]{0.80\textwidth}
        \caption{Unfulfilled P1 criterion (Spotify)}
        \label{fig:p1_spotify}
        \includegraphics[width=\linewidth]{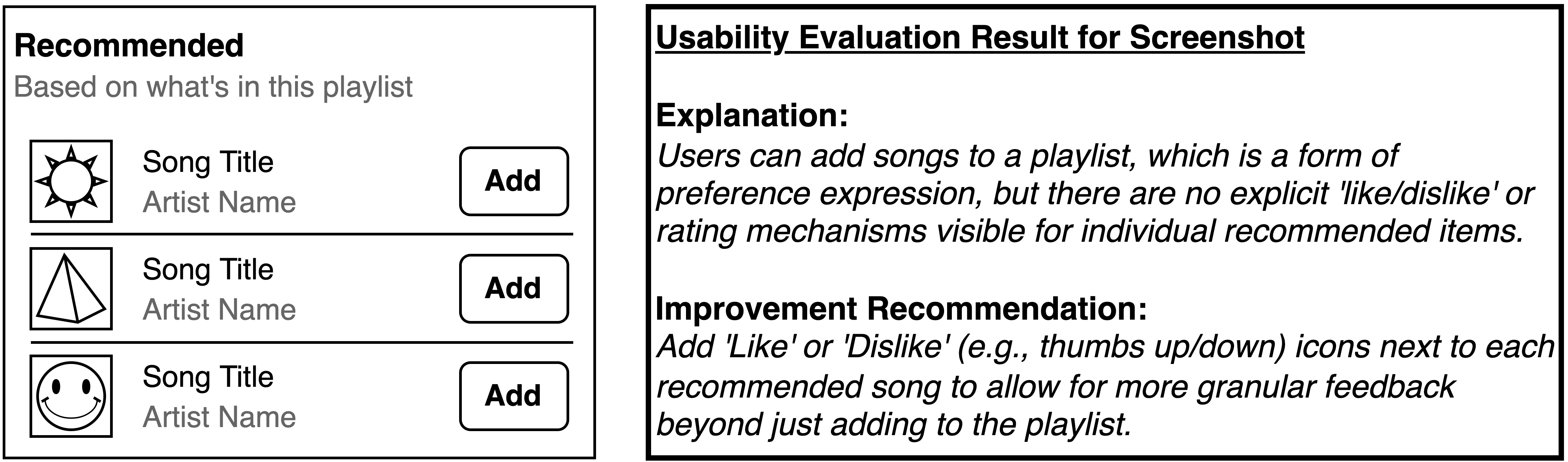}
    \end{subfigure}
    \hfill
    \begin{subfigure}[b]{0.80\textwidth}
        \caption{Fulfilled P1 criterion (Amazon)}
        \label{fig:p1_amazon}
        \includegraphics[width=\linewidth]{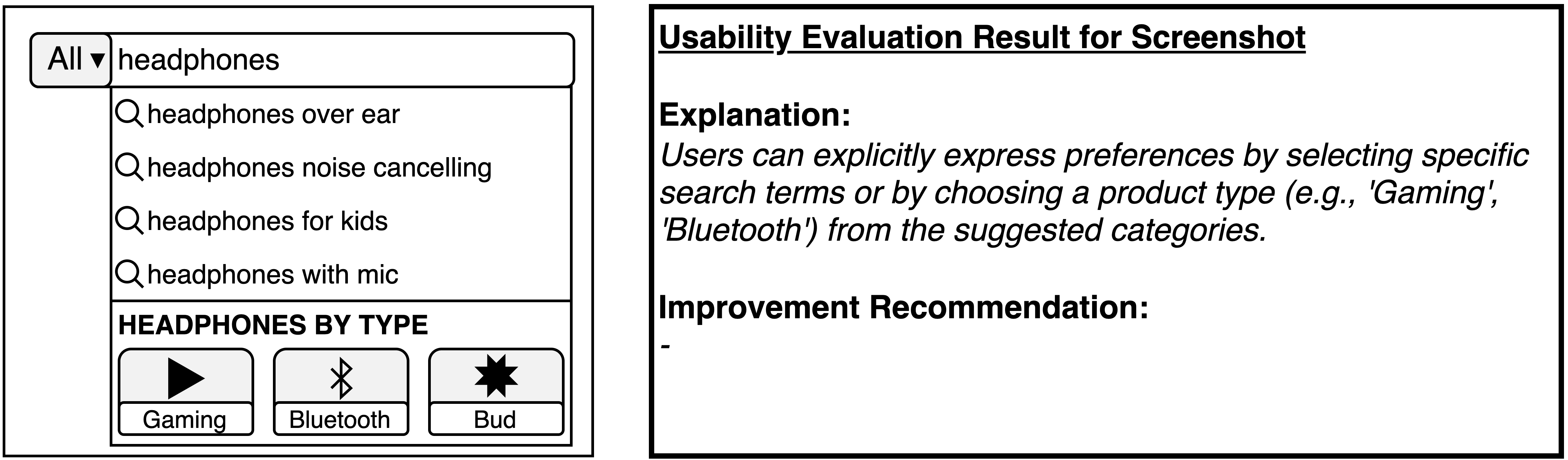}
    \end{subfigure}
    \caption{Example usability analysis result for the \textit{explicit preference expression (P1)} criterion, including explanations and improvement suggestions. The sketches illustrate the analyzed UIs.}
    \label{fig:p1}
    \vspace{-16pt}
\end{figure}

For the recommender-specific criteria, several cases of unfulfilled usability were identified. Figure~\ref{fig:p1} illustrates example results related to criterion P1, which concerns the ability for users to explicitly express preferences. In the playlist creation scenario on \textit{Spotify}, the LLM judged this criterion as unfulfilled, since users can only add recommended songs without any direct feedback mechanism, such as liking or disliking. In contrast, \textit{Amazon} was evaluated as fulfilling the criterion. Users could actively apply clickable filters to indicate item type preferences.

The explanation provided for \textit{Spotify} is particularly nuanced and insightful. While the LLM acknowledges that adding songs to a playlist can be seen as a form of preference expression, it argues that this action alone may not sufficiently satisfy the criterion. This reasoning is persuasive as users might skip a recommendation for various reasons, such as disliking the artist or simply finding the song unsuitable for the current playlist, which is not explicitly communicated to the system. The suggested improvement, to provide more fine-grained feedback options, is both reasonable and actionable.

\begin{figure}[h]
    \centering
    \begin{subfigure}[b]{0.80\textwidth}
        \caption{Unfulfilled R4 criterion (Google News).}
        \label{fig:r4_googlenews}
        \includegraphics[width=\linewidth]{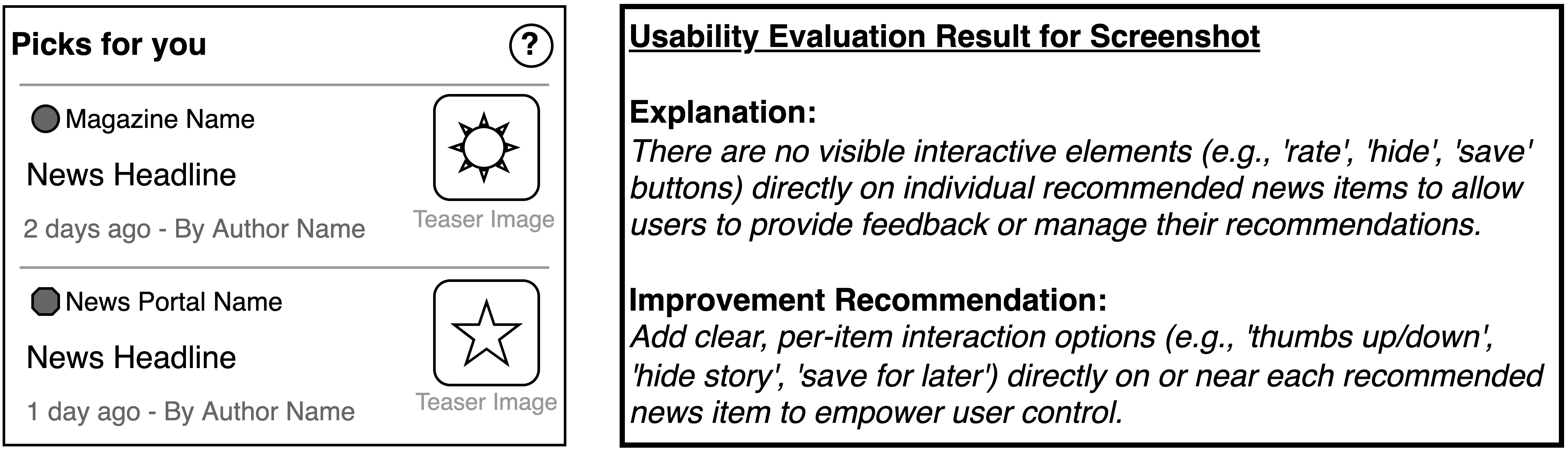}
    \end{subfigure}
    \hfill
    \begin{subfigure}[b]{0.80\textwidth}
        \caption{Fulfilled R4 criterion (YouTube).}
        \label{fig:r4_youtube}
        \includegraphics[width=\linewidth]{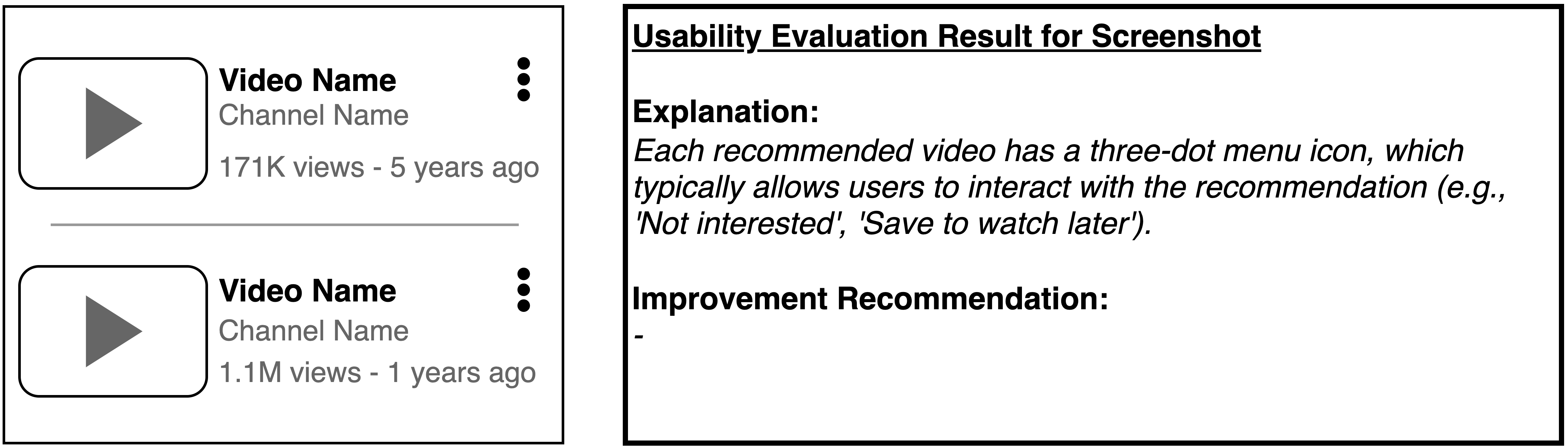}
    \end{subfigure}
    \caption{Example usability analysis result for the \textit{explicit interactiveness of recommended items (R4)} criterion, including explanations and improvement suggestions. The sketches illustrate the analyzed UIs.}
    \label{fig:r4}
    \vspace{-16pt}
\end{figure}

In the \textit{recommendation presentation} scenario, several criteria were considered unfulfilled. Figure~\ref{fig:r4} presents example results for criterion R4, which concerns the ability of users to interact with recommended items. For \textit{Google News}, the LLM noted the absence of visible interaction options and suggested improvements. In contrast, \textit{YouTube} was evaluated more positively, as its recommendation cards include an accessible interaction menu (``\textit{three-dot}'' menu).

These examples highlight both the usefulness and limitations of our approach. While the LLM was able to identify relevant usability gaps, it also operated solely on static screenshots. As a result, interactive elements that appear only on hover or during user interaction, such as the hidden menu in \textit{Google News}, may remain unrecognized. Nevertheless, the LLM's suggestion remains valid, as for a new or inexperienced user, the lack of visible affordances can be a barrier to effective interaction and may justify more prominent cues.

Another notable observation is that the LLM judged none of the platforms as fulfilling criterion R3, which concerns providing explanations for why items are recommended. However, a closer review of the platforms and LLM explanations reveals a more nuanced picture. While many platforms indeed lack explicit explanations, some offer at least high-level contextual hints. For instance, \textit{Spotify} displays recommended playlists with labels such as ``\textit{Brand new music from artists you love}'', which is a high-level explanation for the recommendation. This suggests that the criterion could benefit from further refinement to distinguish between vague contextual hints and explicit, personalized explanations.

\section{Discussion}\label{sec:discussion} 
\vspace{-2pt}
Our results suggest that LLM-based usability analysis can provide useful, low-effort insights into the strengths and weaknesses of recommender interfaces. The generated explanations and improvement suggestions were generally accurate, understandable, and relevant, which indicates the potential for such tools to support development, especially in early-stage prototyping and iterative design. While LLMs cannot replace expert evaluations, they could act as assistive tools in a human-in-the-loop setting, reducing manual effort and accelerating the identification of usability issues.

Building on these findings, several research challenges emerge:

\vspace{-4pt}
\paragraph{Prioritization of Issues.}
The number of identified usability issues of a recommender system can be large and trigger significant effort for developers to evaluate them and prioritize their fixes. The currently used binary fulfillment decisions do not capture the severity of issues, which limits the possibilities for ranking. Using more nuanced assessments using severity ratings could support issue prioritization and make the analysis more actionable by highlighting the most critical usability gaps. 

\vspace{-4pt}
\paragraph{Prompting Design and Context.}
Prompt design can influence the analysis results. Our current template evaluates multiple criteria simultaneously (see Figure~\ref{fig:user_prompt}), which reduces inference costs but sometimes may overlook issues. Using separate prompts for each criterion could improve effectiveness.

Another opportunity for improvement is the refinement of the context description and clarifying the intended meaning of each criterion. In some cases, the LLM struggled to interpret the criteria consistently, for example, determining what level of detail qualifies as an explanation (R3). Making this clearer by adding more detailed descriptions or examples could lead to more accurate and robust results.

\vspace{-4pt}
\paragraph{Dynamic UI Behavior.}
A boundary of our current approach is the reliance on static screenshots, which prevents the model from recognizing dynamic interface elements that appear \textit{on mouse hover}\footnote{see \url{https://www.w3schools.com/howto/howto_css_display_element_hover.asp} for an \textit{on mouse hover} example.} or click. Providing a video of recorded interactions as context could show the complete interface behavior and overcome this limitation. Beyond that, LLM-based agents that directly interact with interfaces could be an even more elaborate approach to also automate the data collection aspect.

\vspace{-4pt}
\paragraph{Validation Against Expert Assessments.}
Systematic comparison with expert evaluations is needed to assess the reliability and practical value of LLM-based usability analysis. While early studies report promising alignment with expert judgments on general usability criteria~\cite{zhong2025synthetic,lubos2025towards}, broader comparisons, particularly in recommender-specific contexts, are needed to validate this.

\section{Conclusions}\label{sec:conclusions} 
\vspace{-2pt}
In this paper, we explored an LLM-based approach to usability analysis of recommender user interfaces. We applied the method to ten publicly available platforms to assess whether the identified issues were plausible, clearly explained, and accompanied by meaningful improvement suggestions. Our findings demonstrate the potential of multimodal LLMs to support low-effort usability evaluation of recommender interfaces, particularly during early-stage design. Also, the findings highlight different areas for improvement, particularly in handling dynamic interface elements and generating more nuanced, context-aware judgments. Future work will focus on improving prompt strategies and contextual understanding, and validating LLM-generated assessments against expert evaluations.


\section*{Declaration on Generative AI}
During the preparation of this work, the authors used \textit{ChatGPT} and \textit{Grammarly} in order to: \textit{Grammar and spelling check}, \textit{Paraphrase and reword}. After using these tools, the authors reviewed and edited the content as needed and take full responsibility for the publication’s content.

\vspace{-4pt}
\bibliography{references}

\end{document}